\documentclass[aps,pra,superscriptaddress,twocolumn]{revtex4}

\usepackage{graphicx} 
\usepackage{amsmath}
 
\newcommand{\beq}{\begin{equation}}
\newcommand{\enq}{\end{equation}}

\begin{document}

\title{Cooper problem in a lattice}
\author{J.-P. Martikainen}
\email{jpjmarti@nordita.org}
\affiliation{Nordita, 106 91 Stockholm, Sweden}
\date{\today}
\begin{abstract}
Cooper problem for interacting fermions
is solved in a lattice. It is found that the binding energy of
the Cooper problem can behave qualitatively differently
from the gap parameter of the BCS theory and that
pairs of non-zero center of mass momentum are favored
in systems with unequal Fermi energies.
\end{abstract}
\maketitle

\section{Introduction}
\label{sec:intro}
Cooper found~\cite{Cooper1956a} out that in the presence of the quiescent
Fermi sea attractively interacting electrons can form a bound state
even if their total energy is larger than zero. In the absence
of Fermi sea such bound state does not exist in a three-dimensional
system in a continuum although such a bound state may appear 
in quasi two-dimensional systems~\cite{Petrov2000b}.
The possibility a bound state indicates instability
and and is traditionally thought of as an indicator of the
instability towards the formation of a superfluid. 
However, the proper theoretical description 
of the superfluid itself involves
a more refined many-body theory than the one used in the Cooper problem.
The purpose of this report is to solve
this classic Cooper problem in a lattice for a polarized system.

After formulating the Cooper problem in a lattice 
several questions are addressed. First, how does the binding energy
behave and is it always qualitatively similar 
to the gap parameter 
in the BCS theory~\cite{Bardeen1957a}? Second, what happens
to the  bound state in systems with unequal Fermi surfaces?
Third, can bound states of non-zero center of mass be favorable
in systems with unequal Fermi surfaces? This
question is of interest since in the BCS theory one expects
modulated, FFLO-type~\cite{Fulde1964a,Larkin1964a}, 
order parameters in systems
with mismatched Fermi surfaces. 
Since optical lattices for ultracold atoms can be made anisotropic
and spin-dependent, the problem is framed and solved 
in a rather general setting with spin-dependent lattice potentials.

Previously some aspects of the two-body problem in a lattice 
have been investigated in the absence of Fermi 
seas in Refs.~\cite{Fedichev2004a,Wouters2006a,Nygaard2008a}.
Dimensionality effects in the Cooper problem were discussed
by Esebbag {\it al.}~\cite{Esebbag1992}.

\section{Formalism}
We assume a two-component system and we label the spin states
by $\sigma=\{\uparrow,\downarrow\}$. If the system is composed
of ultracold neutral fermionic atoms, these "spin'' states
would correspond to either different atoms or
different hyperfine levels of the same isotope.
We further assume that atoms are at zero temperature and in 
a cubic lattice which is deep enough so that only the lowest band must
be considered. Also, since the lattice is deep
it is enough to consider only the leading order nearest
neighbor tunneling processes with tunneling strengths
${\bf t}_{\sigma}$ where bold-face indicates that the
tunneling strengths are represented as vectors and that
the tunneling strength can be different in different directions.

At zero temperature fermions on different spin states
can interact via  $s$-wave interaction and we take the
interaction between 
unequal fermions located at lattice sites ${\bf x}_\downarrow$ and 
${\bf x}_\uparrow$
to be $g\delta({\bf x}_\downarrow-{\bf x}_\uparrow)$. 
Two-body wavefunction for the atoms is then a solution
of the Schr\"{o}dinger equation
\beq
\left[-\sum_\sigma{\bf t}_\sigma\cdot\nabla_\sigma^2
+g\delta({\bf x}_\downarrow-{\bf x}_\uparrow)
\right]\psi({\bf x}_\downarrow,{\bf x}_\uparrow)=
E\psi({\bf x}_\downarrow,{\bf x}_\uparrow).
\nonumber
\enq
Here ${\bf t}\cdot\nabla^2$ denotes the discrete
kinetic energy operator and it acts as
\beq
-{\bf t}\cdot\nabla^2\psi({\bf x})
=-\sum_{\alpha} t_\alpha\left[\psi({\bf x}_\alpha+d) 
-2\psi({\bf x}_\alpha)+\psi({\bf x}_\alpha-d)
\right],\nonumber
\enq
where $d=1$ is the lattice constant
and $\alpha\in\{x,y,z\}$.
Since the interaction only depends on the relative coordinate 
it is useful to write the problem in terms of the wave-function for
the relative coordinate ${\bf x}={\bf x}_\downarrow-{\bf x}_\uparrow$. 
The center of mass and relative motion no longer separate in a lattice, 
but interactions do not couple different center of mass states
to each other.
We write the wavefunction as
\beq
\psi({\bf x}_\downarrow,{\bf x}_\uparrow)=
\Pi_{\alpha\in \{x,y,z\}} e^{iK_\alpha\left(c_\alpha x_{\downarrow,\alpha}+(1-c_\alpha) x_{\uparrow,\alpha}\right)}
\psi({\bf r}),
\enq
where ${\bf c}$ is yet to be determined coefficient related with
the center of mass coordinate in a lattice. Since the center of mass
and the relative coordinate do not separate, the algebra
is somewhat more complicated
than in a continuum, but nevertheless straightforward.
It turns out that the coefficients $c_\alpha$ are determined by the equation
\beq
\frac{t_{\downarrow,\alpha}}{t_{\uparrow,\alpha}}=
\frac{\sin\left(K_\alpha(1-c_\alpha)\right)}{\sin\left(K_\alpha c_\alpha\right)}.
\label{eq_cm}
\enq
If the lattice is spin independent the left hand side is equal to one
and $c_\alpha=1/2$. This corresponds to the usual center of mass
transformation for equal mass atoms. If $\uparrow$-component is very heavy, 
the left hand side becomes very large and the coefficient 
$c_\alpha\rightarrow 0$, which again conforms to our intuition about 
the center of mass coordinate. However,
in case of general spin dependent lattice potentials,
coefficients must be solved using Eq.~(\ref{eq_cm}).

The Schr\"{o}dinger equation for the relative coordinate then becomes
\begin{eqnarray}
&&\left[\sum_\alpha{\bf E}({\bf K})_\alpha \nabla_\alpha^2
+2\left(t_{\downarrow,\alpha}+t_{\uparrow,\alpha}+{\bf E}({\bf K})_\alpha\right)+
\right.\nonumber\\
&+&\left.g\delta({\bf x})\right]\psi({\bf x})=E\psi({\bf x}),
\end{eqnarray}
where 
\beq
{\bf E}({\bf K})_\alpha=
-t_{\downarrow,\alpha}\cos(K_\alpha c_\alpha)-t_{\uparrow,\alpha}\cos(K_\alpha(1-c_\alpha)).
\nonumber
\enq
By expanding the relative coordinate wave function
\beq
\psi({\bf x})=\frac{1}{Nd^3}\sum_{{\bf q}} \psi_{{\bf q}}e^{i{\bf q}\cdot {\bf x}},
\enq
where $N$ is the number of lattice sites, we find 
\begin{eqnarray}
&&\frac{1}{N}\sum_{{\bf q}}
\sum_\alpha\left\{\left[2{\bf E}({\bf K})_{\alpha}\left(\cos({\bf q}_\alpha)-1\right)
\right.\right.\nonumber\\
&+&\left.\left.
2\left(t_{\downarrow,\alpha}+t_{\uparrow,\alpha}+{\bf E}({\bf K})_{\alpha}\right)
\right]
-E\right\}\psi_{{\bf q}}e^{i{\bf q}\cdot {\bf x}}
\nonumber\\
&=&\frac{-g}{N}\sum_{{\bf q}} \psi_{{\bf q}}e^{i{\bf q}\cdot {\bf x}}\delta({\bf x}).
\end{eqnarray}
By multiplying with $e^{-i{\bf k}\cdot {\bf r}}$, summing over
lattice sites, and defining $\alpha=\sum_{{\bf q}} \psi_{{\bf q}}$,
we find an equation 
\beq
\frac{-1}{g}=\frac{1}{N}\sum_{{\bf q}} \frac{1}{f({\bf K},{\bf q})-E},
\label{eq_boundstate}
\enq
where
\begin{eqnarray}
f({\bf K},{\bf q})&=&\sum_\alpha
2{\bf E}({\bf K})_{\alpha}\left(\cos({\bf q}_\alpha)-1\right)
\nonumber\\
&+&2\left(t_{\downarrow,\alpha}+t_{\uparrow,\alpha}+{\bf E}({\bf K})_{\alpha}\right).
\nonumber
\end{eqnarray}

The crucial ingredient of the Cooper problem is the presence of
the occupied Fermi seas. Due to Pauli blocking the two-body wavefunction
cannot have amplitudes for those states that are already occupied.
This restricts the sum-over the first Brillouin zone
$\sum_{{\bf q}}$ into just a sum over allowed states $\sum_{{\bf q}}'$.
If the Fermi energies are $\epsilon_{F,\sigma}$ 
and the free dispersions are 
$\epsilon({\bf q})_{\sigma}=\sum_\alpha 2t_{\sigma,\alpha}(1-\cos({\bf q}_\alpha))$,
then
in practice this means that one should only include those modes
${\bf q}$
that satisfy
$\epsilon({\bf q}+{\bf c}{\bf K})_{\downarrow}>\epsilon_{F,\downarrow}$
and
$\epsilon(-{\bf q}+({\bf 1-c}){\bf K})_{\uparrow}>\epsilon_{F,\uparrow}$
simultaneously. 

By using Eq.~(\ref{eq_boundstate}) one can solve for the 
pair energy $E$ and investigate whether solutions 
with energies less than $\epsilon_{F,\downarrow}+\epsilon_{F,\uparrow}$
exist. For this reason it is helpful to write
$E=\epsilon_{F,\downarrow}+\epsilon_{F,\uparrow}-\Delta$ so that $\Delta>0$
for bound states. Due to the anisotropy and 
generally complicated structure of the Fermi surfaces, integrations
are performed numerically.

\section{Binding energy in symmetric lattices}

\begin{figure}
\includegraphics[width=0.90\columnwidth]{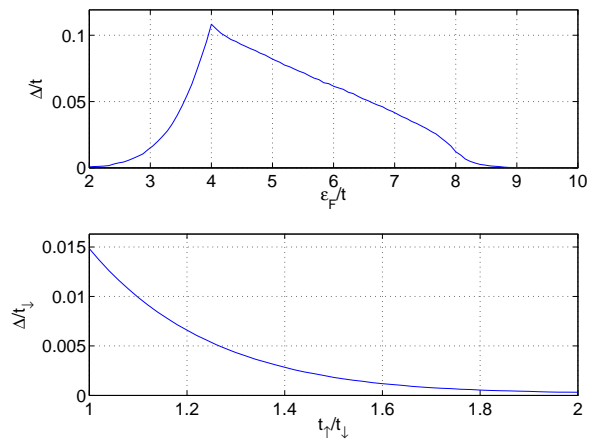} 
\caption{Binding energy in a symmetric lattice with
$g=-3t_x$ as a function of Fermi energy (top) and effective
mass ratio (bottom) with $\mu_\sigma=3t_\sigma$.
}
\label{fig:sweepmu}
\end{figure}

Fig.~\ref{fig:sweepmu} shows the binding energy as a function
of Fermi energy for a system where both components see
an equal and  symmetric lattice as well as
a system where the effective masses of the fermions
are different. In the first case, the binding energy first rises quickly
with the Fermi energy, but at $\epsilon_F=4t$ the topology of the Fermi surface
changes from closed to open. For higher values of the Fermi energy
the binding energy decreases monotonically. Due to the
particle-hole symmetry the gap parameter of the BCS theory is symmetric
with respect to $\epsilon_F=6t$ 
(which corresponds to half-filling and maximum gap parameter 
in the BCS theory).
Such symmetry is absent in the Cooper problem. The behavior at small
Fermi energies, which correspond to low filling fractions,
is similar to the usual free space Cooper problem.
Fig.~\ref{fig:sweepmu}  also shows the binding energy for the case
when fermions have different effective masses i.e. when
$t_\uparrow\neq t_\downarrow$. As the $\uparrow$-component becomes heavier
the binding energy is reduced. This behavior is qualitatively similar
to the behavior of the BCS gap parameter and is to a good accuracy
exponential in the mass ratio.

In free space the solution to the Cooper problem
predicts a binding energy $\Delta\sim \exp(-2/\lambda)$
while the BCS theory predicts a
gap parameter $\Delta_{BCS}\sim \exp(-1/\lambda)$, where
$\lambda=n(\epsilon_F)|g|$ and $n(\epsilon_F)$ is the density
of states at the Fermi-energy. Therefore, for a similar coupling strength
the binding energy in the Cooper problem is $\Delta\sim\Delta_{BCS}^2$,
which is much smaller than $\Delta_{BCS}$ in the weak coupling regime.
In a lattice, $\Delta$ is also much smaller than $\Delta_{BCS}$, but generally
the relation between these two quantities is more complicated.
As the Fig.~\ref{fig:sweepmu} demonstrates, the binding energy of
the Cooper problem, is more sensitive to the structure of the
Fermi-surfaces than the BCS gap parameter.

\section{Binding energy and dimensional crossovers}
\begin{figure}
\includegraphics[width=0.90\columnwidth]{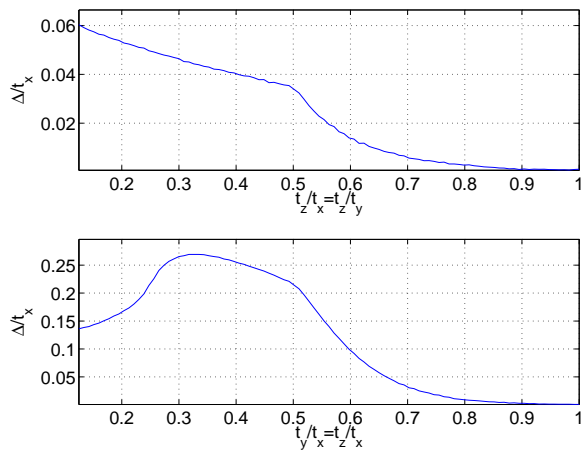} 
\caption{Binding energy as a function of tunneling
strength anisotropy for transitions into quasi-2D (top)
and quasi-1D (bottom) situations.
We choose $\epsilon_{F,\downarrow}=\epsilon_{F,\uparrow}=2t_x$
and $g=-3t_x$.
}
\label{fig:dimcrossover}
\end{figure}

If one makes the lattice anisotropic, one can 
effectively change the dimensionality of the system.
In Fig.~\ref{fig:dimcrossover} demonstrates 
the effect of such changes in two cases. On the
top part one changes the tunneling strength along
$z$-direction while keeping other tunneling strengths the same.
For small values of $t_z$ the lattice is deep in the $z$-direction
and the system is effectively a set of quasi-two-dimensional
systems. As $t_z/t_x$ increases to $\mu/4$ the structure of the 
Fermi surfaces change from  open (cylinder shaped) Fermi-surfaces
into a closed (cigar shaped) ones. This change is reflected as a kink
in the binding energy of the Cooper problem.

The bottom part of Fig.~\ref{fig:dimcrossover}
shows the binding energy for a system where
$t_y$ and $t_z$ are varied, but kept equal to one another.
For small values of $t_y$ and $t_z$ the system is
effectively one-dimensional and Fermi surfaces are disconnected.
Then the Fermi surface is composed of two sheets with
no intersecting points.
At $t_y=t_z=\mu/8$, the Fermi surfaces become
connected, but are still open. The surfaces become connected
so that the corners of the previously disconnected sheets
merge first.
As the tunneling strengths in $y$- and $z$-directions increase further,
at $t_y=t_z=\mu/4$, the Fermi surfaces
become closed and resemble the Fermi surfaces in a continuum
in that sense. 
These changes in the Fermi surface topology
are again reflected as kinks in the binding energy.
Also in this case the binding energy 
is qualitatively different from the BCS gap parameter, which
decreases monotonically as $t_y/t_x=t_z/t_x$ increases.
In the one-dimensional limit fluctuations are important and
the simple BCS mean-field theory is unreliable. It is interesting that
the two-body Cooper problem avoids some qualitative problems
faced by the BCS theory.

\section{Bound states of non-zero center of mass momentum}
In a two-component fermionic gas the atom numbers
of different components can be independently controlled and such
strongly interacting polarized fermion gases
have been recently 
studied experimentally~\cite{Partridge2006a,Zwierlein2006a,Shin2006a}. 
in a harmonic trap.
Studies of polarized fermionic gases have revealed
the possibility of phase separation~\cite{Bedaque2003a} as well as
the possibility of FFLO type order parameters which break
the translational symmetry~\cite{Sheehy2007a}. 
Due to nesting these modulated order parameters
are expected to be more prominent in lattices~\cite{Koponen2007a}
and in systems with reduced dimensionality~\cite{Machida1984a,Parish2007b}.

In the BCS theory at $T=0$ the non-zero order parameter
can exist for Fermi energy differences of order
$\delta\epsilon_F=\epsilon_{F,\uparrow}-\epsilon_{F,\downarrow}
\sim \sqrt{2}\Delta_{BCS}(\delta\epsilon_F=0)$
~\cite{Clogston1962a}.
We have verified that a similar conclusion applies for the existence of the
bound state in the Cooper problem in the lattice. 
I.e. the bound state exists if 
$\delta\epsilon_F$ is (roughly) less than
the binding energy $\Delta$ at $\delta\epsilon_F=0$.
 
In an unpolarized system chemical potentials are the same and 
the Fermi surfaces are also the same. Then
the binding energy of the Cooper problem is maximized
at zero center of mass momentum due
to the reduced density of available low energy states
for non-zero pair momentum. However, when the system is polarized
the Fermi surfaces
are different and this argument is not necessarily valid anymore.
Then non-zero pair momentum might be favorable just like
FFLO-type states can appear in the BCS theory.

\begin{figure}
\includegraphics[width=0.90\columnwidth]{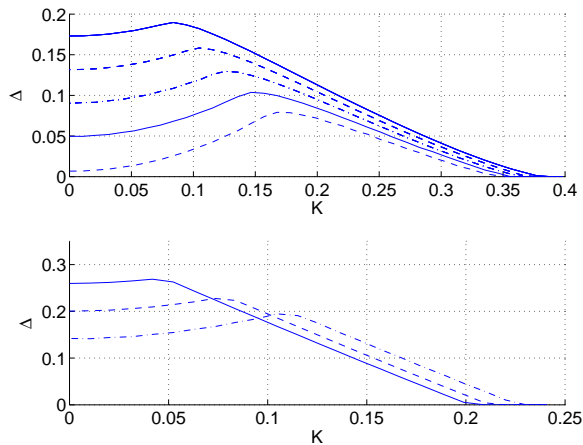} 
\caption{Binding energy as a function of pair center
of mass momentum (along $x$-direction) 
for few different Fermi energy differences $\delta\epsilon_F$.
Top figure: the three-dimensional lattice was symmetric, 
${\bar \epsilon_{F}}=\left(\epsilon_{F,\uparrow}+\epsilon_{F,\downarrow}\right)/2
=4t_x$, $g=-4t_x$, and (from top to bottom)
$\delta\epsilon_F/{\bar \epsilon_{F}}=\{0.04, 0.05, 0.06, 0.07, 0.08\}$.
Bottom figure: lattice was quasi one-dimensional with $t_y=t_z=t_x/10$,
${\bar \epsilon_{F}}=3t_x$,
 $g=-4t_x$, and (from top to bottom)
$\delta\epsilon_F/{\bar \epsilon_{F}}=\{0.03, 0.05, 0.07\}$.
}
\label{fig:sweepK}
\end{figure}

As Fig.~\ref{fig:sweepK} demonstrates, for systems with unequal Fermi 
surfaces the binding energy can indeed be maximized at non-zero center
of mass momentum. If the Fermi energy difference is too large
the bound state does not exist, but for Fermi energy differences
which are less than about 
$\Delta(\delta\epsilon_F=0)$
the possibility of non-zero center of mass pair must be
taken into account. Fig.~\ref{fig:sweepK}
also shows the binding energy in a quasi one-dimensional
lattice. The binding
energy of the non-zero center of mass pair is typically 
much larger in the one-dimensional system than a three-dimensional one. 
In the figure the binding energies in 
three-dimensional and one-dimensional systems are roughly
similar in magnitude only because the average Fermi energy
was lower in the quasi one-dimensional problem.
One-dimensional system is qualitatively different from
the three-dimensional one in that as Fermi energy difference increases
the existence of bound states persists to larger
center of mass momenta.  This is reminiscent of the BCS theory in
one-dimensional systems where there is no upper critical polarization
above which the order parameter disappears.                      

\section{Conclusions}
In summary, we solved the Cooper problem in a lattice and found
that the behavior of the binding energy of the Cooper problem
is qualitatively different from the behavior of the
gap parameter of the BCS theory. Also, for systems with
different Fermi energies, if the instability exists
it is towards the formation of pairs with non-zero center
of mass momentum. In  three dimensions at zero temperature
the BCS theory for a polarized system predicts
FFLO ordering only above certain non-zero Fermi energy 
difference~\cite{Sheehy2007a,Koponen2007a}.
For smaller Fermi energy differences phase
separation is expected to occur~\cite{Bedaque2003a}.
In contrast, the Cooper problem with mismatched Fermi surfaces
always predicts instability towards
non-zero center of mass pairs.

Two new interesting articles~\cite{Piil2008a,Valiente2008a} 
discussing the two-body 
scattering in an one-dimensional  lattice in the absence of Fermi 
seas have appeared after the submission of this report.

\bibliographystyle{apsrev}

\end{document}